\documentclass[aps,prl,twocolumn,showpacs,superscriptaddress,floatfix]{revtex4}
\usepackage{graphicx}

\newcommand{\rt}{\rightarrow}

\newcommand{\be}{\begin{equation}}
\newcommand{\ee}{\end{equation}}
\newcommand{\bea}{\begin{eqnarray}}
\newcommand{\eea}{\end{eqnarray}}

\begin{document}

\title{Violation of Porod law in a freely cooling granular gas in one dimension} 
\author{Mahendra Shinde}
\affiliation{
Department of Physics, Indian Institute of Technology Bombay, 
Powai, Mumbai 400076, India}
\author{Dibyendu Das}
\affiliation{
Department of Physics, Indian Institute of Technology Bombay, 
Powai, Mumbai 400076, India}
\author{R. Rajesh}
\affiliation{
Institute of Mathematical Sciences, CIT Campus, Taramani, Chennai-600013,
India}
\date{\today}

\begin{abstract}
We study a model of freely cooling inelastic granular gas in one
dimension, with a restitution coefficient which approaches the elastic
limit below a relative velocity scale $\delta$. While at
early times ($t \ll \delta^{-1}$) the gas behaves as a completely
inelastic sticky gas conforming to predictions of earlier studies, at
late times ($t \gg \delta^{-1}$) it exhibits a new fluctuation
dominated phase ordering state.  We find distinct scaling behavior for
the (i) density distribution function, (ii) occupied and empty gap
distribution functions, (iii) the density structure function and (iv)
the velocity structure function, as compared to the completely
inelastic sticky gas. The spatial structure functions (iii) and (iv)
violate the Porod law. Within a mean-field approximation, the exponents
describing the structure functions are related to those describing the
spatial gap distribution functions.
\end{abstract}

\pacs{47.70.Nd, 64.75.+g, 05.70.Ln, 45.70.Mg}

\maketitle

Flowing granular media exhibit varied physical phenomena 
\cite{nagel,kadanoff}. 
A simple, well studied model that captures many features of flowing
granular systems is a gas of particles undergoing inelastic collisions
\cite{kadanoff,carnevale,goldhirsch,frachebourg,bennaim1,bennaim2,
cattuto,puglisi,puri}. The system may be externally driven or allowed
to cool freely.  
However, real granular gases have {\it
elastic} collisions when relative velocity of particles approach zero
\cite{raman}.  Thus, a realistic model of cooling granular gas
should have a relative velocity dependent restitution coefficient
\cite{raman}. In this paper we focus on such a model.

In general, a system freely relaxing to an ordered state has a macroscopic
length scale ${\cal L}(t)$ increasing with time $t$ \cite{bray}. In addition,
for usual phase ordering systems, the presence of a dominant ${\cal L}(t)$
results in a robust scaling law called the Porod law \cite{bray,porod}. For
scalar order parameters, the Porod law states that the scaled structure
function $S/{\cal L}^d \sim (k{\cal L})^{-{\theta}}$ for large $k{\cal L}$,
with $\theta = 2$ in one dimension. 
Contrary to this clean scenario, in certain driven systems 
\cite{das,manoj,nagar,ramaswamy}, an unusual phase
ordering was observed. These systems have a macroscopic coarsening
length scale ${\cal L}$ but the domain lengths have a power-law
distribution with a large negative power. This leads to a violation of
the Porod law with $\theta\neq 2$. We shall refer to such systems as
fluctuation-dominated phase ordering (FDPO) systems.  We demonstrate below 
that a coarsening granular gas too has such an unusual FDPO state.

The issue of an inelastic gas showing coarsening and phase ordering
has been addressed in many earlier publications on the subject
\cite{frachebourg,bennaim1,bennaim2,cattuto,puri, lattice1,lattice2}.
It is now well-known that both the freely
cooling inelastic gas ($0<r<1$) and the sticky gas ($r=0$)
undergo coarsening with a
growing length scale ${\cal L}(t) \sim t^{1/z}$, with $z=3/2$ in one
dimension \cite{carnevale,frachebourg,bennaim1}.  The sticky gas
problem in one dimension can be solved exactly and is known to be
equivalent to the inviscid Burgers equation
\cite{frachebourg,kida}. From this equivalence, the structure
functions of the sticky gas can be inferred to obey the Porod law
\cite{kida}.

Numerical studies have tried to relate the behavior of the inelastic
gas, to the sticky gas.  It was shown that at large times the
decay of total energy of the inelastic gas is identical to the sticky
gas \cite{bennaim1}.  Moreover, other quantities like ${\cal L}(t)$
and velocity distribution function have the same scaling form as that
of the sticky gas. This suggested that for any deviation from the
elastic limit, the large time scaling behavior crosses over to that
of the sticky gas and that the underlying continuum equation is the
inviscid Burgers equation \cite{bennaim1}.

In this paper, we show that if a granular gas is modeled as a gas of
particles having a velocity dependent restitution coefficient, 
interesting new physics appear. In
the late time regime, the equivalence with the sticky gas breaks down
and the system exhibits a FDPO state. For our granular gas model, 
velocity dependence of the restitution coefficient is chosen to be:  
\be
r = (1 - r_0) {\rm exp}\left(-|v_{\rm rel}/\delta|^{\sigma}\right) + r_0.  
\label{r}
\ee 
For relative velocity $v_{\rm rel} \ll \delta$,  $r \rt 1$, and for 
$v_{\rm rel} \gg
\delta$,  $r \rt r_0<1$ , mimicking
the experimental scenario \cite{raman}. The parameter $\sigma$
determines how sharply the crossover from $r_0$ to $1$ happens across
the crossover scale $\delta$. While  experiments 
\cite{raman} suggest a wide range of values for $\sigma$, kinetic theory
studies estimate $\sigma = 1/5$ \cite{kinetic_sigma}.  We
note that  taking first the limit $\sigma \rt \infty$ and then
$\delta \rt 0$, this model becomes the same as the model 
studied in Ref.~ \cite{bennaim1}.

Our main result is that the cooling granular gas has a time scale $t_1
\sim \delta^{-1}$, such that the density distribution function, and
various {\it spatial} distribution functions show a complete change of
behavior across it. Yet at the same time, the total energy $E(t)$
decay as $\sim t^{-{2/3}}$ throughout, without any signature of change
across the timescale $t_1$. For $t \ll t_1$, we find that the granular
gas behaves as a sticky gas (as in
\cite{bennaim1}). But, for late times $t_1 \ll t
\ll t_2$ (where $t_2 \sim \delta^{-3}$), the phase ordering is
distinct from the sticky gas. In particular, the density-density and
velocity-velocity structure functions show violation of the Porod
law. We note that the scale $t_2$, beyond which all collisions are
elastic and $E(t)$ stops decreasing, is  easily
understood in terms of the velocity scaling law 
\cite{bennaim1} -- $v(t_2) \sim t_2^{-{1/3}} \sim \delta$ implying
$t_2 \sim \delta^{-3}$. On the contrary, the interesting scale $t_1$
that we find is much smaller.
   
We now define our model more precisely. We consider $N$ point
particles of equal mass on a ring of length $L$. Initially, the particles
are distributed randomly in space with their velocities drawn from a
normal distribution. The particles undergo
inelastic, momentum conserving collisions such that when
two particles with velocity $u_i$ and $u_j$ collide, the final velocities
$u_i^{\prime}$ and $u_j^{\prime}$ are given by: 
\be 
u_{i,j}^{\prime} =
u_{i,j} \left(\frac{1-r}{2} \right) + u_{j,i} \left(\frac{1+r}{2}\right).
\ee 
We define coarse-grained densities and velocities for the granular gas
as follows \cite{puglisi}.  At any point of time the system is divided
into $N$ equally sized spatial boxes. The total number of particles in
the $i$th ($i=1,2,\dots,N$) box defines the mass density $\rho_i$. The
velocity $v_i$ is defined as the sum of the velocities of the
particles in box $i$.  For the sticky gas it
suffices to talk about distributions of masses and velocities of
individual particles.
\begin{figure}
\includegraphics[width=8.0cm]{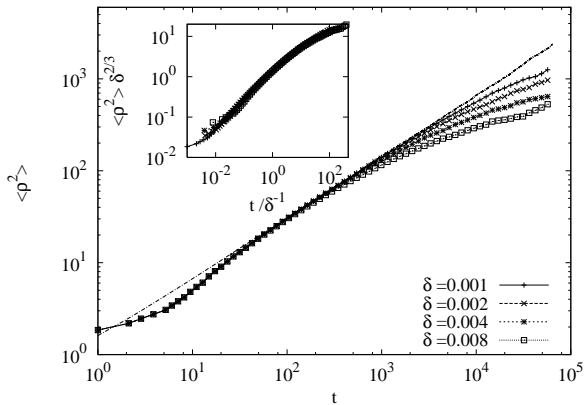}
\caption{\label{fig1} The variation of $\langle {\rho}^2 \rangle$ with 
time $t$ is shown for different values of $\delta$  when  $\sigma=3.0$.
The dashed 
line corresponds to  the sticky gas. Inset: The 
curves collapse when $t$ is scaled by $\delta^{-1}$.
}
\end{figure}

We have done an event driven molecular dynamics simulation for system
sizes $L$, ranging from $20000 - 50000$ (in units of inter particle
spacing). The particle density is set to $1$ throughout.  The system
was evolved up to time $t=32000 - 64000$ (in units of initial mean
collision time). For these times ${\cal L}(t) \ll L$.  Simulations
were done for different $\sigma$ (namely $3$, $4$, $5$, $10$ and
$\infty$), $\delta$ (namely $0.001-0.01$), and $r_0$ (namely $0.1$,
$0.5$ and $0.8$) values. 
There was no qualitative difference found for
the various sets of these parameter values. Hence, we choose a specific
set for the data presented below -- $r_0 = 0.5$, $\sigma = \infty$,
$\delta = 0.01$ (unless otherwise mentioned).  Whenever there is
a quantitative dependence on $\sigma$, we mention it.
  
The existence of the time scale $t_1$ can be seen by examining
$\langle \rho^2 \rangle$, where $\rho$ is the density.
In Fig.~\ref{fig1} we show the time dependence of $\langle
\rho^2 \rangle$ for systems with same $\sigma = 3$ but different
$\delta$ values. 
At early times $\langle \rho^2 \rangle \sim t^{2/3}$ as for the
sticky gas. The departure from
the sticky gas curve happens at a timescale $t_1$ which increases with
decreasing  $\delta$. The curves collapse when $t$ is scaled by $\delta^{-1}$
(see inset).  We have checked that this dependence is independent of $\sigma$,
i.e.,
$t_1 \sim \delta^{-1}$ for all $\sigma$.

Currently we do not have a deeper understanding of the
timescale $t_1$. Intriguingly, we find no signature of 
$t_1$ in the decay of the total energy $E(t) \sim
t^{-{2/3}}$, which is related to the second
moment of the velocity distribution. In this paper, we concentrate on the  density distribution
and various {\it spatial} distribution functions. The
spatial distributions include the empty and occupied gap distribution
functions, and the density and velocity structure functions.  For the
early time regime $t \ll t_1$,  these
quantities are  numerically equivalent to   the corresponding
quantities of the sticky gas \cite{mahendra}. In this 
paper, we focus on the late time regime $t \gg t_1$, where
there is deviation from the sticky gas.
\begin{figure}
\includegraphics[width=8.0cm]{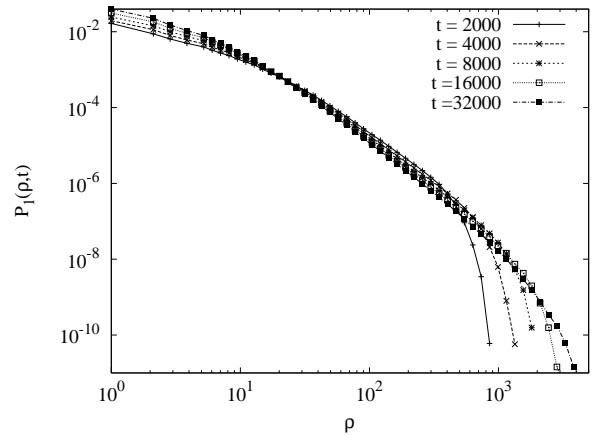}
\caption{\label{fig2} The variation of $P_1(\rho,t)$ with 
$\rho$ is shown for different times. 
}
\end{figure}

Let $P_{1}(\rho,t)$ be the probability that a 
box has mass density $\rho$ at time $t$.  In Fig.~\ref{fig2}, the variation of
$P_{1}(\rho,t)$ with $\rho$ is shown for different times. 
For small $\rho$ ($\rho<10$), $P_1(\rho,t)$ increases
with time to a nonzero constant. For intermediate $\rho$ the slope
increases to a constant. The cutoff increases to infinity with
time. From these features, we conclude that  when $t\gg1$, $P_1(\rho,t)$ 
approaches a
non-zero time independent power law distribution, i.e. 
\be
\lim_{t \gg t_1} \lim_{L \rt \infty} P_1(\rho,t) \sim \rho^{-\gamma_1}.
\label{prho}
\ee 
Conservation of density ($\langle \rho \rangle = 1$)
implies that $\gamma_1 > 2$. Consistent with this, we find that the
power law for the curves at various times extrapolate asymptotically
to $\gamma_1 \simeq 2.83$.  The cutoff ${\rho}_{max}(t)$ scales as
${\cal{L}}(t) \sim t^{\beta}$ with $\beta = 2/3$. 
There are strong corrections to scaling, as is clear
from the change in apparent slope for different times. Hence, it is
not possible to obtain data collapse by scaling unless one measures
$P_{1}(\rho,t)$ for even larger densities and times. The value of $\gamma_1$
depends on the value of $\sigma$. For $\sigma$ varying from $3$ to $\infty$,
$\gamma_1$ varies from $2.30$ to $2.83$.

The data of Fig.~\ref{fig2} for our granular gas implies a scenario in
which small density clusters do not get depleted from the system. This
is in  contrast to the sticky gas, where the probability of the
mass clusters , $P_{1s}(m,t) \sim m^{-1/2}t^{-1/3}$ for $m \ll
t^{2/3}$ \cite{frachebourg} decays to zero with time.  
The coarsening
process in the sticky gas is one of pure aggregation, transferring
mass from  smaller to larger mass scales. On the
other hand, our model at late times effectively shows a combined
aggregation and fragmentation dynamics. This simply comes as the
``elastic'' collisional break-ups at late times (and small velocities)
compete with the aggregation due to inelasticity. The effective rates
are such that mass loss and gain at small scales are balanced out.
\begin{figure}
\includegraphics[width=8.0cm]{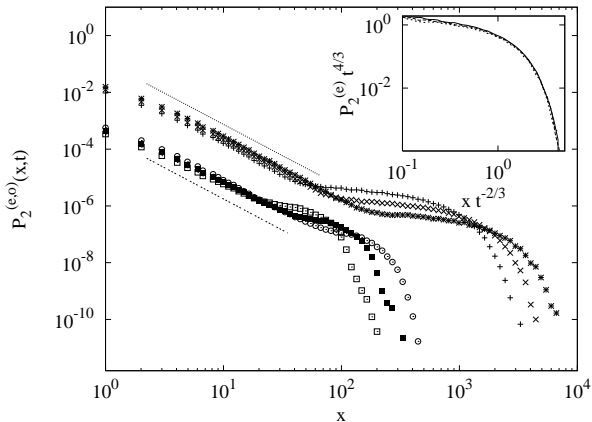}
\caption{ \label{fig3}
The variation of $P_2^{(e,o)}(x,t)$ with $x$ is shown for times
$t=16000, 32000, 64000$ (the larger times have larger cutoffs).
$P_2^{(o)}(x,t)$ is shifted downwards for clarity.
The straight lines have power $-2.2 $.
Inset: The plateaus of $P_2^{(e)}(x,t)$ have the scaling form
$t^{-4/3} f(x t^{-2/3})$.
}
\end{figure}

Further support to the above picture comes from the inter particle and
inter hole gap distribution functions. Let $P_2^{(e,o)}(x,t)$ be the
probability of finding a gap of exactly $x$ empty(e) or occupied(o)
boxes.  The variation of $P_2^{(e,o)}(x,t)$ with $x$ for different
times is shown in Fig.~\ref{fig3}. For small masses, they decay as a
power law with power ${\gamma_2} \simeq -2.2$.  For large masses,
$P_2^{(o)} (x,t)$ has a plateau eventually cutoff at scales $x \sim
{\cal L}(t)$. The shape of the plateau and cutoff is reminiscent of
the the number distribution of gaps in the sticky gas which have the
form $N_{2s}(x,t) = {t^{-4/3}}f_{2s}(x/{{\cal L}(t)})$ with $f_{2s}(z)
\rt 1$ for $z \ll 1$ \cite{frachebourg}. The large $x$ of of
$P_2^{(o)}(x,t)$ scales exactly as $N_{2s}(x,t)$ (see inset of
Fig.~\ref{fig3}). Since $\int dx N_{2s} \sim t^{-{2/3}}$, the area
under the plateau will eventually go to zero.  So while the big gaps
in our model scale as that of the sticky gas, the more important 
fact is that it has predominantly small gaps associated with 
the following power law:  
\be 
\lim_{t \gg t_1} \lim_{L \rt \infty} P_2^{(e,o)}(x,t) \sim x^{-\gamma_2}.
\label{pgap}
\ee 
We have checked that the exponent $\gamma_2$ has no discernible
dependence on $\sigma$.

The inter particle gap distribution has a bearing on the 
density-density and velocity-velocity correlation functions. In
particular, the large power $\gamma_2$ suggests that the ordering
process will be affected by the abundance of smaller gaps and Porod
law could be violated. For the sticky gas, numerical results \cite{kida,mahendra}
confirm that  Porod law holds for both  
density-density and velocity-velocity correlation functions. We now
discuss the case of our granular gas and show that it is different.
\begin{figure}
\includegraphics[width=8.0cm]{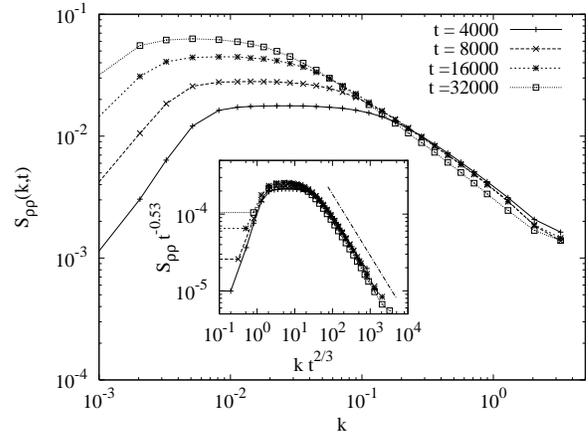}
\caption{\label{fig4} 
The variation of $S_{\rho\rho}(k,t)$ with $k$ is shown
for different times. The inset shows the data scaled 
as in Eq.~(\ref{Sk}).
The straight line has power $-0.80$.}
\end{figure}

Let $C_{\rho\rho}(x,t) = \langle \rho_i(t)
\rho_{i+x}(t) \rangle$ be the equal time density-density correlation function. 
The structure function
$S_{\rho\rho}(k,t)$ is the Fourier transform of
$C_{\rho\rho}(x,t)$. Similarly we define the equal time velocity-velocity
correlation function as $C_{vv}(x,t) = \langle
v_i(t) v_{i+x}(t) \rangle$, with its corresponding structure function
$S_{vv}(k,t)$.
We  show that the structure functions can be expressed in
terms of the gap distribution through a mean field approximation.
 $C_{\rho\rho}(x)$ is approximately
equal to density square times the probability there is a non-zero density
at $x$ given there is a non-zero density at $0$. Thus (setting $\rho=1$), 
$C_{\rho\rho}(x)
\approx \sum_{n=0}^{\infty} p_{2n}(x)$, where $p_{2n}(x)$ is the probability
of having exactly $n$ empty gaps between $0$ and $x$ with $0$ and $x$ being
occupied. 
Let $\tilde{C}_{\rho\rho}(s)$ and $\tilde{P}_2^{(o,e)}(s)$ be the Laplace 
Transforms of $C_{\rho\rho}(x)$and $P_2^{(o,e)}(x)$ respectively. Approximating joint distributions
 by products of individual  distributions \cite{satya}, we obtain
\bea
\tilde{p}_{2n} (s) &=& 
\frac{
\left[1-\tilde{P}_2^{(o)}
\right]^2 \tilde{P}_2^{(e)} \left[ \tilde{P}_2^{(o)}
\tilde{P}_2^{(e)}
\right]^{n-1}}
{\langle x \rangle_o s^2} 
, ~n\geq 1, \label{eq:1}\\
\tilde{p}_{0}(s)  &=& \frac{1}{s}-\frac{1-\tilde{P}_2^{(o)}}{s^2 
\langle x \rangle_o} \label{eq:2},
\eea
where $\langle x \rangle_o = \int x P_2^{(o)}(x) dx$.
Eqs.~(\ref{eq:1}) and (\ref{eq:2}) give
\be
\tilde{C}_{\rho\rho}(s) = \frac{1}{s} - \frac{[1-\tilde{P}_2^{(o)}(s)]
[1-\tilde{P}_2^{(e)}(s)]}
{\langle x \rangle_o s^2 [1-\tilde{P}_2^{(e)}(s) \tilde{P}_2^{(o)}(s)]}.
\ee
The correlation function $C_{vv}(x)$ is  exactly the same
 except for a factor of  $\langle v^2 \rangle$. 
Thus,
$\tilde{C}_{vv}(s) \sim v_t^2 \tilde{C}_{\rho\rho} (s)$.
\begin{figure}
\includegraphics[width=8.0cm]{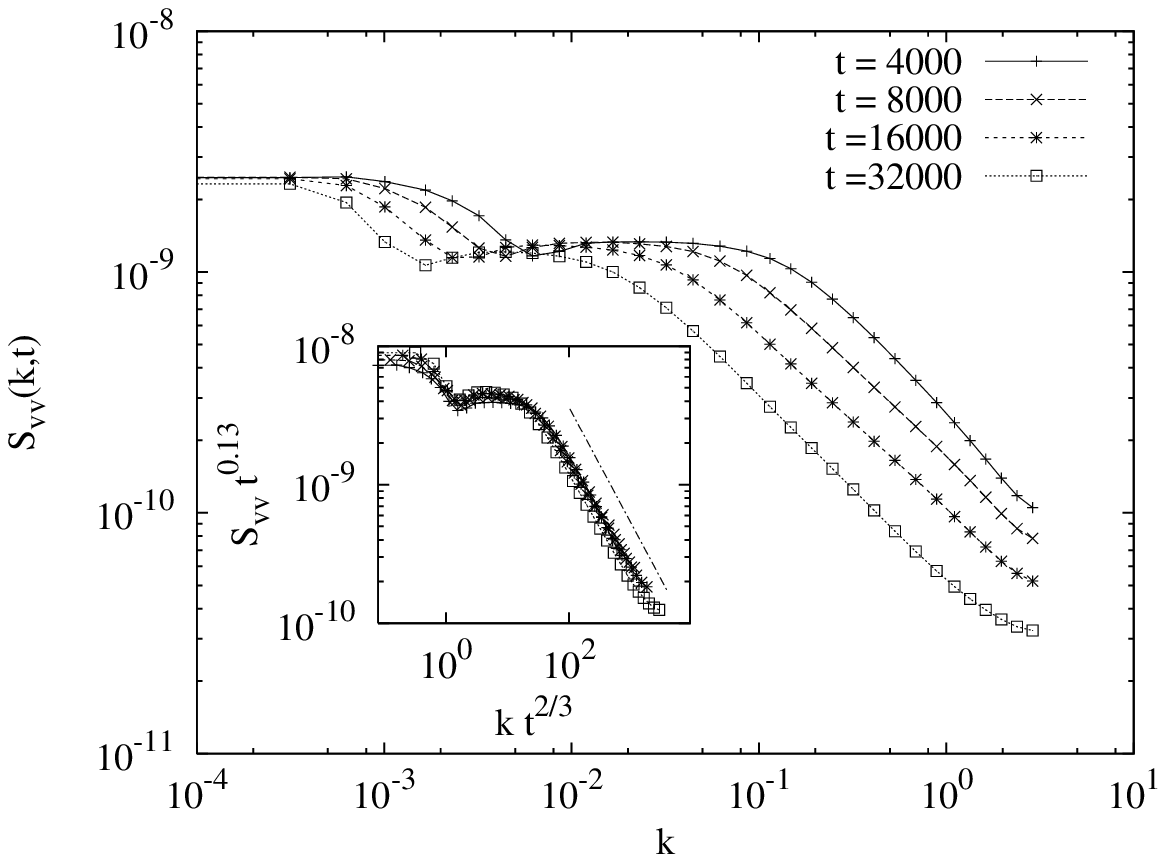}
\caption{\label{fig5} 
The variation of $S_{vv}(k,t)$ with $k$ is shown
for different times. The inset shows the data scaled 
as in Eq.~(\ref{Sv}).
The straight line has power $-0.80$.}
\end{figure}

Now, $1-\tilde{P}_2^{(e,o)}(s) \sim s^{\gamma_2-1}$. Also, $\langle x
\rangle_o \sim {\cal L}^0$ since $\gamma_2 > 2$. This  implies
that $\tilde{C}_{\rho\rho}(s) \sim s^{\gamma_2-3}$, where we have ignored the
$s^{-1}$ which contributes to $\delta(k)$ in the Fourier transform. Thus, 
$S_{\rho\rho}(k)$ will have the scaling form
\be
S_{\rho\rho}(k) = {\cal L}^{3-\gamma_2} f_3(k{\cal L}),
\label{Sk}
\ee
where $f_3(x) \sim x^{-\theta_{\rho}}$ for $x\gg 1$, with
$\theta_{\rho}= 3-\gamma_2$. 
Using $\langle v^2 \rangle  \sim {\cal L}^{-1}$ \cite{bennaim1}, we obtain
\be
S_{vv}(k) = {\cal L}^{2-\gamma_2} f_4(k{\cal L}),
\label{Sv}
\ee
where $f_4(x) \sim x^{-\theta_v}$ for $x\gg 1$ with
$\theta_v=3-\gamma_2$. 
Since $\gamma_2 \approx 2.20$,
$\theta_v=\theta_{\rho} \approx 0.80$ which  
is very different from $2.0$ seen in Porod law
in one dimension. 

The mean field approximation gives a good description of the actual
problem.  In Fig.~\ref{fig4} we show the variation of
$S_{\rho\rho}(k,t)$ with $k$.  The data collapses onto one curve when
scaled as in Eq.~(\ref{Sk}) [see inset of Fig.~\ref{fig4}].  The
scaling function $f_3(z)$ varies as $z^{-0.8}$ for large $z$. In
Fig.~\ref{fig5}, the variation of $S_{vv}(k,t)$ with $k$ is shown.
Again a good collapse is obtained when the data is scaled as in
Eq.~(\ref{Sv}) such that $f_4(z) \sim z^{-0.8}$ for large $z$.  
We found that the exponents $\theta_{\rho}$ and $\theta_v$ have 
no dependence on $\sigma$. 

To summarize, we studied a model of  freely cooling granular
gas with velocity dependent restitution coefficient. We showed the
existence of a time scale $t_1\sim \delta^{-1}$ beyond which the system
deviates from the sticky gas behavior. The effective dynamics in this
regime is one of
aggregation and fragmentation.
As a result, the spatial distribution functions 
change their forms drastically. We found new power law
exponents associated with the density distribution function
and  empty and occupied gap distribution
functions. The two-point spatial correlation functions
violate Porod law and the structure function decay exponent is $
\simeq 0.8$ instead of the usual value $2$. This deviation and the
existence of power laws in the one-point functions indicate that the
phase ordering is dominated by large scale fluctuations.
We hope that 
experiments on cooling granular gases in quasi one-dimension would 
find the deviation in Porod law predicted by us.

We  thank M. Barma, A. Sain and B. Chakraborty for 
discussions.  
D.D. was supported by grant no. $3404-2$ of ``Indo-French Center  
(IFCPAR)/
(CEFIPRA)''.

\end{document}